\documentclass[useAMS,usenatbib]{mn2e}
\usepackage[dvips]{graphicx}

\title[Arecibo timing observations of 17 Pulsars]
      {Arecibo timing and single-pulse observations of 17 pulsars}
\author[D.~J.~Champion et al.]
{D.~J.~Champion$^{1}$,\thanks{Email: champion@jb.man.ac.uk} D.~R.~Lorimer$^{1}$, M.~A.~McLaughlin$^{1}$, K.~M.~Xilouris$^{2}$,
\newauthor
Z.~Arzoumanian$^{3}$, P.~C.~C.~Freire$^{4}$, A.~N.~Lommen$^{5}$, J.~M.~Cordes$^{6}$ and F.~Camilo$^{7}$\\
$^{1}$ University of Manchester, Jodrell Bank Observatory, Macclesfield, Cheshire, SK11 9DL, UK\\ 
$^{2}$ University of Arizona, Steward Observatory, 933 N. Cherry Av. Bldg 65, Tucson, AZ 85721, USA\\
$^{3}$ USRA/EUD, NASA Goddard Space Flight Center, Code 662, Greenbelt, MD 20771, USA\\
$^{4}$ National Astronomy and Ionosphere Center, Arecibo Observatory, HC3 Box 53995, PR 00612, USA\\ 
$^{5}$ Department of Physics and Astronomy, Franklin and Marshall College, Box 3003, Lancaster, PA 17604, USA\\
$^{6}$ Department of Astronomy and Space Sciences, Cornell University, Ithaca, NY 14853, USA\\
$^{7}$ Columbia Astrophysics Laboratory, Columbia University, 550 West 120th Street, New York, NY 10027, USA
}

\begin{document}

\date{\today}

\pagerange{\pageref{firstpage}--\pageref{lastpage}} \pubyear{2004}

\maketitle

\label{firstpage}

\begin{abstract}
We report on timing and single-pulse observations of 17 pulsars discovered at the Arecibo observatory. The highlights of our sample are the recycled pulsars J1829+2456, J1944+0907 and the drifting subpulses observed in PSR~J0815+0939. For the double neutron star binary J1829+2456, in addition to improving upon our existing measurement of relativistic periastron advance, we have now measured the pulsar's spin period derivative. This new result sets an upper limit on the transverse speed of 120 km~s$^{-1}$ and a lower limit on the characteristic age of 12.4~Gyr. From our measurement of proper motion of the isolated 5.2-ms pulsar J1944+0907, we infer a transverse speed of $188 \pm 65$~ km~s$^{-1}$. This is higher than that of any other isolated millisecond pulsar. An estimate of the speed, using interstellar scintillation, of $235 \pm 45$~km~s$^{-1}$ indicates that the scattering medium along the line of sight is non-uniform. We discuss the drifting subpulses detected from three pulsars in the sample, in particular the remarkable drifting subpulse properties of the 645-ms pulsar J0815+0939. Drifting is observed in all four components of the pulse profile, with the sense of drift varying among the different components. This unusual `bi-drifting'' behaviour challenges standard explanations of the drifting subpulse phenomenon.
\end{abstract}


\section{Introduction}
Since the first systematic pulsar survey using the Arecibo telescope \citep{ht74, ht75b, ht75a} which discovered the original double neutron star binary B1913+16 \citep{ht75b}, there have been a number of searches carried out with this sensitive system (e.g. \citealt{sstd86, nft95}). Timing of the pulsars discovered in these searches has provided a wealth of information, including the first evidence for the existence of gravitational radiation \citep{tw82, wt84} and the first extra-solar planetary system \citep{wf92}.

Between 1994 and 1998 the telescope underwent a major upgrade, during which the receivers were parked on the meridian but observations could still be made as the sky drifted past. Pulsar surveys conducted in this drift-scan mode have resulted in the discovery of 130 pulsars \citep{rdk+95, cnst96new, rtj+96, lzb+00, lma+04, mac+02, lxf+05}.

In this paper we report on timing and single-pulse observations of 17 pulsars discovered at Arecibo. Nine of the pulsars were discovered in our recent drift-scan searches \citep{mlc+04, clm+04}, while the remainder were found in earlier surveys and have not been subjected to a follow-up campaign of timing observations \citep{rtj+96, zcwl96, nic99, mac00}. We discuss the timing results including the proper motion of millisecond pulsar (MSP) J1944+0907 and an updated timing solution for the relativistic binary J1829+2456. We then describe the drifting subpulses seen in three of the pulsars, including J0815+0939. This unique pulsar shows drifting in all four components of its unusual pulse profile, with the subpulses in one component drifting in a different sense to the others.

\section{Observations and Analysis}
The observations presented here were carried out between February 1999 and March 2005 using the Arecibo telescope. Most of the observations used the 430-MHz Gregorian dome receiver, which has an average gain of 11~K~Jy$^{-1}$ and an average system temperature of 45 K, in combination with the Penn State Pulsar Machine (PSPM), a 128-channel analogue filterbank spectrometer \citep{cad97a}. During the early stages of timing when the periods of the pulsars were not well determined, the PSPM was used in search mode. The PSPM samples incoming voltages from the telescope every $80 \mu$s over a bandwidth of 7.68~MHz. These data were then dedispersed by delaying successive channels in time corresponding to the nominal dispersion measure (DM) of each pulsar to produce time series which were subsequently folded to produce integrated pulse profiles. To confirm that the period was the fundamental rather than a harmonic, the data was folded at twice the period and the profile was checked. Once the period of each pulsar had been determined accurately, the timing mode of the PSPM was used. In timing mode the data in each frequency channel are folded on-line. The 128 individual profiles can then be dedispersed and summed to produce a single profile.

The first stage of timing analysis is to produce a high signal-to-noise ratio (S/N) profile to use as a template. Initially this was done by shifting the profile so the peak was at a phase of 0.5 before summing the profiles. The time of arrival (TOA) of each profile was then calculated by convolving the profile with the template and adding this offset to the start time of the observation. These TOAs were then analysed with the TEMPO software package\footnote{http://pulsar.princeton.edu/tempo} to produce a preliminary phase-connected solution (where every rotation of the pulsar is accounted for) over the time-span of the observations. In most cases a simple spin down model with five free parameters (pulse phase, period, period derivative, right ascension, and declination) was used. For PSR~J1944+0907 the proper motion is a significant effect and is included in the fit. The relativistic binary pulsar J1829+2456 requires an additional five classical and two relativistic parameters to produce a phase connected solution. For each pulsar the resulting ephemeris was used to predict the pulse phase of each profile. These were then aligned before they were summed to produce a better template which was subsequently used to generate new TOAs, weighted by S/N, which were analysed using TEMPO. This additional iteration often resulted in significant reduction of the root-mean-square observed minus model timing residual. With the exception of PSR~J1917+0834, a single template was sufficient for the timing analysis. For J1917+0834, due to the short length of the integrations and long period of the pulsar, only a small number of pulses were summed causing the relative strengths of the two peaks in the profile to vary from observation to observation, a template for each mode was required to provide accurate TOAs in the initial stages.

Most of the previously published DM measurements were based on the original observations and so were poorly constrained. In December 2004 the pulsars were observed at 320, 327, 334 and 430~MHz using the PSPM. The TOAs produced were analysed using TEMPO by holding all the, now well constrained, parameters of the ephemeris constant and fitting for DM. This new value of DM was then used to dedisperse the data again and improve the template, TOAs and ultimately the ephemerides.

To calibrate the intensity scale of the pulse profiles, we followed the procedure described by Lorimer, Camilo \& Xilouris (2002)\nocite{lcx02} and scaled each profile by the factor $1000 T/(DG)$ mJy, where $T$ is the total system noise temperature (K), $D$ is the DC offset of the profile (in PSPM counts) and $G$ is the antenna gain (K~Jy$^{-1}$). Due to the design of the Arecibo telescope, both $T$ and $G$ are strong functions of zenith angle. This dependence is described analytically by Perillat (1999)\nocite{per99a}. The total system temperature also includes the background sky temperature contribution \citep{hssw82}. The error is based on the standard deviation of the observed fluxes.

\section{Timing Results}
\begin{figure*}
\includegraphics[angle=0,scale=.65]{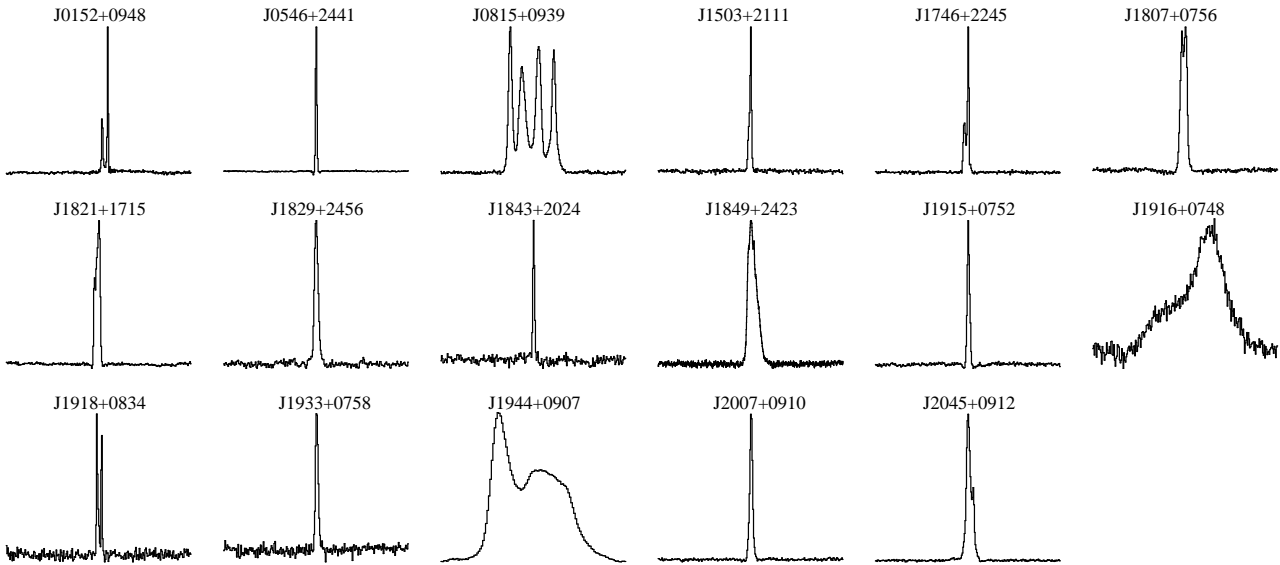}
\caption{Integrated 430-MHz pulse profiles. All profiles show full pulse phase over 256 bins with the exception of PSR~J1944+0907 which has 128 bins. PSR~J1944+0907 is dispersion-smeared over $\sim$4 bins. These profiles are freely available as part of the EPN database.}
\label{Profiles}
\end{figure*}

The basic timing parameters, pulse widths and flux densities are summarised in Table~\ref{ObsPar}. Listed are the pulsar name, position (in right ascension and declination), period $P$, its derivative $\dot{P}$ and epoch, DM, span of observations, number of TOAs and the post-fit weighted rms of the timing residuals. Table~\ref{AddPar} lists the additional timing parameters for the binary pulsar J1829+2456 and the isolated MSP J1944+0907. The 430-MHz profiles derived from a phase-coherent sum over many individual observations are shown in Fig. \ref{Profiles}.

\begin{table*}
\caption{Observed parameters\label{ObsPar}}
\begin{center}
\scriptsize
\begin{tabular}{c l l l r@{.}l c r@{.}l r c r}
\hline
Name & \multicolumn{1}{c}{R.A. (J2000)} & \multicolumn{1}{c}{Decl. (J2000)}         & \multicolumn{1}{c}{$P$} & \multicolumn{2}{c}{$\dot{P}$}    & Epoch & \multicolumn{2}{c}{DM} & \multicolumn{1}{c}{Span} & $N_{\rm TOAs}$ & RMS\\
PSR  & \multicolumn{1}{c}{(h:m:s)}      & \multicolumn{1}{c}{($^{\circ}$:$'$:$''$)} & \multicolumn{1}{c}{(s)} & \multicolumn{2}{c}{$(10^{-15})$} & (MJD) & \multicolumn{2}{c}{(pc cm$^{-3}$)}& \multicolumn{1}{c}{(MJD)}& & $\mu$s\\
\hline
J0152+0948 & 01:52:23.73(5)    & +09:48:10(2)       & 2.74664729014(10)     & 1&700(4)      & 52858 & 21&87(2)  &52597--53004&26&350\\
J0546+2441 & 05:46:28.76(3)    & +24:41:21(6)       & 2.84385038524(6)      & 7&65(3)       & 52914 & 73&81(3)  &52707--53119&29&706\\
J0815+0939 & 08:15:08.776(9)   & +09:39:50.7(6)     & 0.64516119106(5)      & 0&139(6)      & 52854 & 52&667(9) &52600--52796&53&270\\
J1503+2111 & 15:03:54.60(4)    & +21:11:09.3(5)     & 3.31400150122(12)     & 0&14(2)       & 52857 & 11&75(6)  &52595--53118&12&941\\
J1746+2245 & 17:46:00.84(3)    & +22:45:28.9(8)     & 3.46503778264(6)      & 4&919(7)      & 52765 & 52&0(20)  &52401--53129&19&1240\\
J1807+0756 & 18:07:51.235(2)   & +07:56:43.39(5)    & 0.464300493199(2)     & 0&1294(2)     & 53064 & 89&29(3)  &52765--53362&22&134\\
J1821+1715 & 18:21:13.503(4)   & +17:15:47.06(7)    & 1.366682059422(10)    & 0&8714(11)    & 53064 & 60&47(7)  &52765--53362&23&240\\
J1829+2456 & 18:29:34.6668(2)  & +24:56:18.193(3)   & 0.04100982358117(4)   & 0&0000525(15) & 52887 & 13&8(5)   &52785--53447&50&22\\
J1843+2024 & 18:43:26.18(7)    & +20:24:54.6(12)    & 3.4065386707(4)       & 1&04(4)       & 53064 & 85&3(2)   &52765--53362&19&3829\\
J1849+2423 & 18:49:34.6763(4 ) & +24:23:45.976(12)  & 0.27564149867101(7)   & 0&152582(4)   & 52199 & 62&239(5) &51194--52923&244&409\\
J1915+0752 & 19:15:01.94(2)    & +07:52:09.2(6)     & 2.05831378861(7)      & 0&139(9)      & 53048 & 105&3(3)  &52733--53362&60&2891\\
J1916+0748 & 19:16:51.5(2)     & +07:48:00(5)       & 0.54175211381(17)     & 10&708(17)    & 53048 & 304&0(15) &52733--53362&15&14754\\
J1917+0834 & 19:17:48.853(6)   & +08:34:54.63(14)   & 2.129665364024(4)     & 17&4935(2)    & 52345 & 29&18(6)  &51327--53362&36&608\\
J1933+0758 & 19:33:19.407(2)   & +07:58:07.01(7)    & 0.437454407217(2)     & 0&2183(2)     & 53048 & 165&02(4) &52732--53362&19&171\\
J1944+0907 & 19:44:09.32084(3) & +09:07:23.2388(12) & 0.0051852019041260(3) & 0&00001713(2) & 52913 & 24&34(2)  &52732--53447&27&2.6\\
J2007+0910 & 20:07:58.1031(12) & +09:10:12.93(6)    & 0.4587348258074(6)    & 0&33244(12)   & 52994 & 48&6(2)   &52595--53392&25&116\\
J2045+0912 & 20:45:47.3438(9)  & +09:12:29.16(4)    & 0.3955551117394(5)    & 0&19520(7)    & 52997 & 31&776(4) &52601--53392&25&78\\
\hline
\end{tabular}
\normalsize
\end{center}
Observed parameters of the 17 pulsars derived from the TEMPO timing analysis described in section 2. Figures in parentheses are 1-$\sigma$ uncertainties in the least significant digits. The arrival times from which these ephemerides are derived are freely available online as part of the European Pulsar Network (EPN) database (http://www.jb.man.ac.uk/$\sim$pulsar/Resources/epn).
\end{table*}

\begin{table*}
\caption{Additional parameters for pulsars J1829+2456 and J1944+0907\label{AddPar}}
\begin{center}
\begin{tabular}{lll}
\hline
PSR\\
\hline
J1829+2456 &                                                                       & \\
           & $x$, Projected semi-major axis (lt-sec)                               & 7.238(2)\\
           & $P_b$, Binary period (days)                                           & 1.176027941(16)\\
           & $e$, Eccentricity                                                     & 0.1391412(19)\\
           & $\omega$, Longitude of periastron (deg)                               & 229.92(2)\\
           & T$_0$, Epoch of periastron (MJD)                                      & 52848.579775(3)\\
           & $\dot{\omega}$, Rate of advance of periastron (deg yr$^{-1}$)         & 0.2919(16)\\
           & $\gamma$, Relativistic time-dilation and gravitational redshift (ms)  & 3.6(38)\\
           & Mean TOA uncertainty ($\mu$s)                                         & 33.7\\
J1944+0907 &                                                                       & \\
           & $\mu_\alpha$, Proper motion in R.A. (mas yr$^{-1}$)                   & 12.0(7)\\
	   & $\mu_\delta$, Proper motion in Decl. (mas yr$^{-1}$)		   & $-$18(3)\\
           & Mean TOA uncertainty ($\mu$s)                                         & 3.6\\
\hline
\end{tabular}
\end{center}
Figures in parentheses are 1-$\sigma$ uncertainties in the least significant digits.
\end{table*}

Further observed and derived parameters are summarised in Table~\ref{DerPar}. Listed are the pulse widths at 10\%, $w_{10}$, and 50\%, $w_{50}$, of peak amplitude; the equivalent pulse width (the width of a top hat function with the same integrated flux and peak amplitude), $w_{eq}$; the mean 430-MHz phase-averaged flux density, $S_{430}$; distance $d$ (inferred from the free electron density model of Cordes \& Lazio 2002\nocite{cl02}); Galactic height ($z=d \sin{b}$, where $b$ is the Galactic latitude); 430-MHz luminosity ($L_{430}\equiv S_{430} d^{2}$); spin-down age ($\tau = P/2\dot{P}$); spin-down energy-loss rate ($\dot{E} = 4 \pi^{2} I \dot{P}/P^{3}$, where the neutron star moment of inertia $I$ is taken to be $10^{45}$~g~cm$^{2}$); and inferred surface dipole magnetic field strength ($B = 3.2 \times 10^{19}\sqrt{P\dot{P}}$~G; Lorimer \& Kramer 2005\nocite{lk05}).

\begin{table*}
\caption{Further observed and derived parameters\label{DerPar}}
\begin{center}
\begin{tabular}{l r@{}l r@{}l r@{}l r@{.}l r@{.}l r@{.}l r@{.}l r@{.}l r@{.}l r@{.}l}
\hline
\multicolumn{1}{c}{Name} & \multicolumn{2}{c}{$w_{10}$} & \multicolumn{2}{c}{$w_{50}$} & \multicolumn{2}{c}{$w_{eq}$} & \multicolumn{2}{c}{$S_{430}$} & \multicolumn{2}{c}{$d$}   & \multicolumn{2}{c}{$z$}   & \multicolumn{2}{c}{$L_{430}$}       & \multicolumn{2}{c}{Log $\tau$} & \multicolumn{2}{c}{Log $\dot{E}$}       & \multicolumn{2}{c}{Log $B$} \\
\multicolumn{1}{c}{PSR}  & \multicolumn{2}{c}{(ms)}     & \multicolumn{2}{c}{(ms)}     & \multicolumn{2}{c}{(ms)}     & \multicolumn{2}{c}{(mJy)}     & \multicolumn{2}{c}{(kpc)} & \multicolumn{2}{c}{(kpc)} & \multicolumn{2}{c}{(mJy kpc$^{2}$)} & \multicolumn{2}{c}{(kyr)}      & \multicolumn{2}{c}{(ergs s$^{-1}$)}     & \multicolumn{2}{c}{(G)} \\
\hline
J0152+0948 & 103&      & 11&       & 17&       & 0&91(3)  & 0&9 & $-$0&71 &  0&7     & 4&4     & 30&5     & 12&3      \\
J0546+2441 & 51&       & 28&       & 25&       & 2&65(10) & 2&0 & $-$0&07 & 10&6     & 3&8     & 31&1     & 12&7      \\
J0815+0939 & 183&      & 166&      & 75&       & 3&7(2)   & 2&5 & 0&98    & 23&1     & 4&9     & 31&3     & 11&5      \\
J1503+2111 & 80&       & 26&       & 39&       & 1&3(2)   & 1&0 & 0&88    &  1&3     & 5&6     & 29&2     & 11&8      \\
J1746+2245 & 125&      & 34&       & 50&       & 1&26(7)  & 3&0 & 1&22    & 10&8     & 4&1     & 30&7     & 12&6      \\
J1807+0756 & 27&.3     & 18&.2     & 17&.6     & 2&16(6)  & 3&5 & 0&81    & 26&5     & 4&8     & 31&7     & 11&4      \\
J1821+1715 & 64&       & 45&       & 39&       & 3&7(2)   & 2&8 & 0&70    & 29&0     & 4&4     & 31&1     & 12&0      \\
J1829+2456 & 1&.7      & 0&.8      & 1&.0      & 0&3(1)   & 1&2 & 0&33    &  0&4     & 7&1     & 31&5     &  9&2      \\
J1843+2024 & 29&       & 40&       & 29&       & 0&37(3)  & 4&0 & 0&75    &  5&9     & 4&7     & 30&0     & 12&3      \\
J1849+2423 & 26&.1     & 14&.4     & 14&.6     & 2&3(6)   & 3&2 & 0&63    & 23&6     & 4&5     & 32&5     & 11&3      \\
J1915+0752 & 60&       & 24&       & 28&       & 1&70(10) & 3&8 & $-$0&11 & 24&6     & 5&4     & 29&8     & 11&7      \\
J1916+0748 & 230&      & 80&       & 50&       & 1&14(9)  & 8&1 & $-$0&29 & 74&8     & 2&9     & 33&4     & 12&4      \\
J1917+0834 & 81&       & 68&       & 35&       & 0&44(10) & 1&9 & $-$0&06 &  1&6     & 3&3     & 31&9     & 12&8      \\
J1933+0758 & 14&.6     & 6&.8      & 6&.8      & 0&34(3)  & 6&0 & $-$0&58 & 12&2     & 4&5     & 32&0     & 11&5      \\
J1944+0907 & 2&.5      & 0&.5      & 0&.5      & 3&9(3)   & 1&8 & $-$0&23 & 12&6     & 7&1     & 33&2     &  8&3      \\
J2007+0910 & 16&.6     & 8&.3      & 8&.7      & 1&51(5)  & 2&0 & $-$0&55 &  6&0     & 4&3     & 32&1     & 11&6      \\
J2045+0912 & 25&.5     & 10&.4     & 13&.7     & 3&5(2)   & 1&9 & $-$0&66 & 12&6     & 4&5     & 32&1     & 11&5      \\
\hline
\end{tabular}
\end{center}
Pulse profile widths, 430-MHz flux density measurements and various derived parameters for the 17 pulsars. All distance measurements have statistical error of 25\% as estimated by Cordes \& Lazio (2002), individual errors may be larger. Figures in parentheses are 1-$\sigma$ uncertainties in the least significant digits for the observed parameters. For PSR~J1944+0907, the Shklovskii-corrected  $\dot{P}$ has been used to calculate the luminosity, spin-down age, spin-down energy loss rate and inferred surface dipole magnetic field strength.
\end{table*}

\begin{figure}
\includegraphics[angle=270,scale=.30]{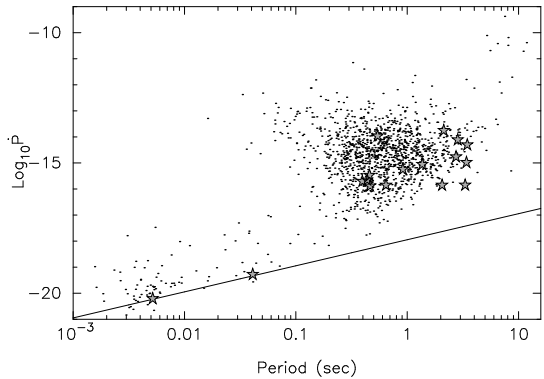}
\caption{The $P-\dot{P}$ diagram showing all currently known pulsars listed in the on-line Australia Telescope National Facility (ATNF) pulsar calatalogue (http://www.atnf.csiro.au/research/pulsar/psrcat). The stars indicate the 17 pulsars in the current sample. Young pulsars are in the top right of the plot while old recycled pulsars are in the bottom left. The line indicates the age of the Universe, $\tau = 14\ \rm{Gyr}$, equivalent to $H_0 = 70\ \rm{km\ s^{-1}\ Mpc^{-1}}$ \citep{yjb+05}. The $\dot{P}$ of PSR~J1944+0907 has been Shklovskii-corrected assuming the DM-inferred distance (see section \ref{shk1944}). The $\dot{P}$ of PSR~J1829+2456, for which there currently no proper motion measurement, has not been Shklovskii-corrected (see section \ref{1829shk}).}
\label{PPdot}
\end{figure}

In Fig.~\ref{PPdot} we show the locations of the 17 pulsars on the period-period derivative ($P$-$\dot{P}$) diagram. As can be seen, the majority of the pulsars occupy the lower end of the `island' of points suggesting that they are relatively old members of the normal population. This is in keeping with previous samples of pulsars drawn from low-frequency surveys (e.g. \citealt{lcx02}) which are less effective at finding young pulsars than surveys at higher frequencies (e.g. \citealt{mlc+01}).

\subsection{PSR~J1829+2456}
\begin{figure}
\includegraphics[angle=0,scale=.45]{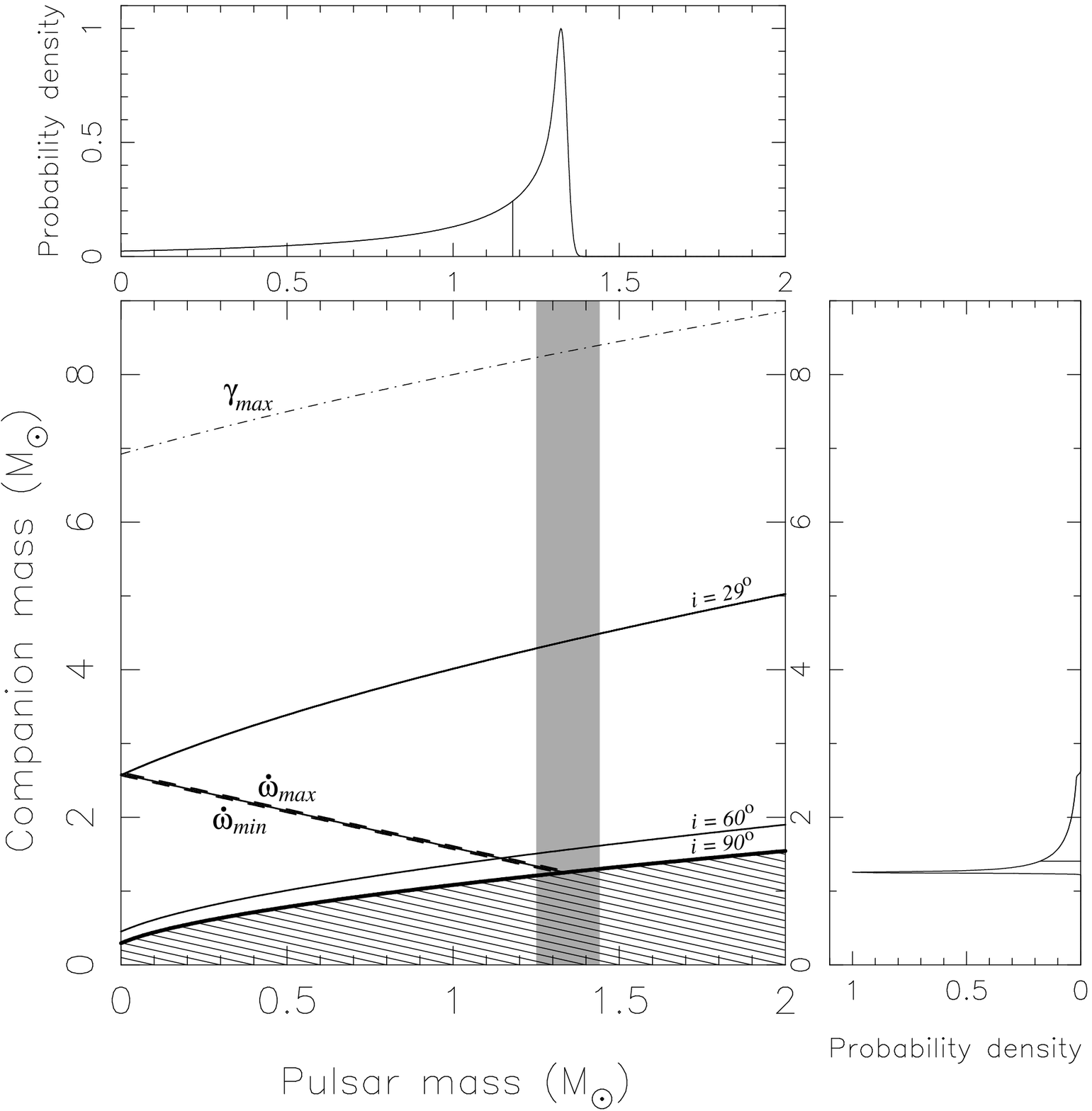}
\caption{Current constraints on the masses of PSR~J1829+2456 and companion. The grey bar shows the mass limits for neutron stars, the upper limit described by Thorsett \& Chakrabarty (1999), and the low measured mass of PSR J0737$-$3039B (Lyne et al.~2004). There is currently only an upper limit for the time-dilation and redshift parameter, $\gamma$. The upper and lower limits of $\dot{\omega}$ are now so close they appear as one thick line. The masses of the system are constrained to be between the two $\dot{\omega}$ limits and above the $i=90^{\circ}$ limit. The marginal figures show the probability density functions for the pulsar and companion masses assuming a random distribution of inclination angles. The lines of constant mass indicate the median of the distributions.}
\label{M-Mfig}
\end{figure}

This 41-ms pulsar is in a mildly eccentric binary system where the companion is most likely a neutron star \citep{clm+04}. The rate of advance of periastron, $\dot{\omega}$, is now measured to a significance of 180-$\sigma$. Interpreting $\dot{\omega}$ as purely relativistic precession implies that the total mass of the binary $M=2.59\pm0.02$~M$_{\odot}$ (see Fig. \ref{M-Mfig}). The lower companion mass limit of 1.26~M$_{\odot}$, and upper limit to the pulsar mass of 1.34~M$_{\odot}$ are given by the constraint that the inclination angle $i\leq 90^{\circ}$. A very conservative upper limit on the companion mass $M_{c}=2.61$~M$_{\odot}$, and a minimum inclination angle $i>29^{\circ}$, can also be inferred. A more likely upper limit of $M_{c}=1.33$~M$_{\odot}$ is obtained by assuming a minimum pulsar mass of 1.26~M$_{\odot}$ (the lowest measured mass of any neutron star, namely PSR J0737$-$3039B; \citealt{lbk+04}). An upper limit of 8~ms can now also be placed on the relativistic time-dilation and gravitational redshift parameter, $\gamma$, but, as shown in Fig.~\ref{M-Mfig}, this is not yet constraining. The determination of these limits is described in more detail in \cite{clm+04}.\nocite{tc99} \nocite{lbk+04}

\label{1829shk}
The extended baseline of our TOAs has also resulted in a measurement of $\dot{P}$ for this pulsar. However this $\dot{P}$ should be treated with caution since any transverse component of proper motion of the system would lead to secular acceleration \citep{Shk70, ctk94} and will contribute to the observed period derivative,

\begin{equation}
\frac{\dot{P}}{P} = \frac{1}{c}\frac{V_{\rm{T}}^{2}}{d} = 2.43\times10^{-21}\rm{s}^{-1}\left(\frac{\it{d}}{\rm{kpc}}\right)\left(\frac{\mu_{\rm{T}}}{\rm{mas\ yr^{-1}}}\right)^{2},
\label{shkequ}
\end{equation}

\noindent where $\dot{P}$ contribution due to the proper motion, $V_{\rm{T}}$ is the transverse speed and $\mu_{\rm{T}}$ is the proper motion. Indeed, assuming the current measurement of $\dot{P}$ is entirely due to proper motion, and adopting the nominal DM-derived distance of 1.2~kpc, the Shklovskii effect places an upper limit of 20~mas~yr$^{-1}$ on the proper motion, corresponding to a transverse speed of $\sim$118 km s$^{-1}$. This transverse speed is consistent with other double neutron star systems so it is possible that the intrinsic $\dot{P}$ is considerably lower. 

However if the $\dot{P}$ of PSR~J1829+2456 is assumed to be intrinsic it places this double neutron star system in an area of the $P$-$\dot{P}$ diagram inhabited by various eccentric binary systems (see Fig. \ref{PPdot}). This $\dot{P}$ implies a characteristic age of at least 12.4 Gyr and a magnetic field strength of $1.5\times 10^{9}$ G. As for other recycled pulsars this large characteristic age suggests that its birth spin period was comparable to its current spin period. Of the other double neutron star systems, only J1518+4904 has a lower $\dot{P}$ \citep{hlk+04}. PSR~J1829+2456 is currently the subject of an extended timing campaign to measure a proper motion and so remove any contribution of the Shklovskii effect from the measurement of $\dot{P}$. Simulated TOAs suggest that a proper motion of 15~mas~yr$^{-1}$ should be measured to $\sim 30\%$ and the expected $\gamma$ (1.3 ms) should be measured to $\sim 40\%$ accuracy within the next year. Simulations also suggest that it would be possible to measure the Shapiro delay with the current observing equipment if the inclination was $\sim90^\circ$.

\subsection{PSR~J1944+0907}
\begin{figure}
\includegraphics[angle=270,scale=.38]{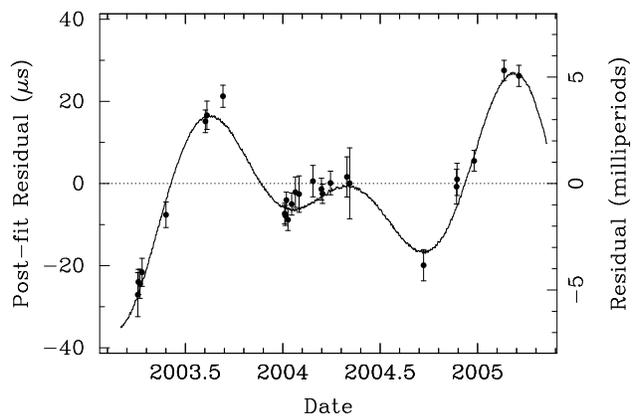}
\caption{The timing residuals for PSR~J1944+0907 before fitting for proper motion. The solid line indicates the fit for a proper motion of 22~mas~yr$^{-1}$.}
\label{propermotion}
\end{figure}

PSR~J1944+0907 is an isolated 5.2-ms pulsar with a broad profile \citep{mac+05}.  Isolated MSPs are rare objects; only 13 are currently known \citep{Lor05}. While binary MSPs are thought to be spun up by the accretion of matter from an evolved binary companion, the formation process for isolated MSPs is presently not clear.  Either they form through a different formation channel (such as accretion-induced collapse), or they ablate their companion with their strong relativistic wind \citep{rst89a}. As shown in Fig.~\ref{propermotion}, the proper motion of this pulsar, $\mu_{\rm{T}}$, is detected very significantly in our timing observations from which we obtain $\mu_{\rm{T}} = 22 \pm 2$~mas~yr$^{-1}$. This would imply a transverse speed $V_{T}=\mu_{\rm{T}} d=188 \pm 65$~km~s$^{-1}$; this is significantly higher than the average for recycled systems ($87 \pm 13$, Hobbs et al. 2005\nocite{hllk05}) and, with the exception of PSR~B1257+12 which has a planetary system, this is by far the highest of all isolated (and indeed most binary) MSPs so far \citep{cnt93, cnt96, nt95, cbl+95, tsb+99, wdk+00, lzb+00, eb01, lma+04, fsk+04, hfs+04}.

\label{shk1944}
The very large transverse speed implied by the proper motion would have a significant effect on the observed $\dot{P}$ as described in \ref{1829shk}. When this Shklovskii effect is taken into account the intrinsic $\dot{P}$ drops to $6.2\times10^{-21}$. This value has been used to calculate the derived parameters listed in Table~\ref{DerPar}.

\begin{figure}
\includegraphics[angle=0,scale=.80]{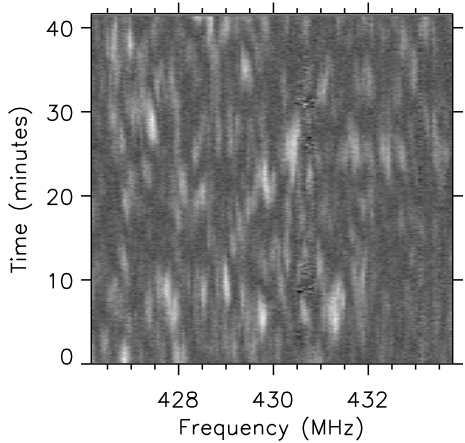}
\caption{Dynamic spectrum (i.e. on-pulse intensity versus time and frequency) for J1944+0907 at 430~MHz. The intensity is indicated by the linear greyscale (black: low intensity, white: high intensity). The spectrum was formed with 10-s integrations using the PSPM, with 128 frequency channels across the 7.68-MHz bandwidth.}
\label{1944fig}
\end{figure}

PSR~J1944+0907 also scintillates strongly at 430~MHz, as shown in Fig. \ref{1944fig}. An auto-correlation analysis of its dynamic spectrum indicates a bandwidth and timescale of $\Delta\nu_d$ = 90~kHz and $\Delta t_d$ = 101~s. The interstellar scintillation derived transverse speed, $V_{ISS}$, is given by
\begin{equation}
V_{ISS} = \frac{A d^{\frac{1}{2}} \Delta \nu_d^{\frac{1}{2}}}{\nu \Delta t_d},
\label{issequ}
\end{equation}
where the constant $A = 2.53\times 10^4$ km~s$^{-1}$ for a uniform Kolmogorov scattering medium, $d$ is the distance in kpc, $\nu$ is frequency in GHz, $\Delta \nu_d$ is scintillation bandwidth in MHz and $\Delta t_d$ is the scintillation timescale in seconds \citep{cr98}. Assuming a distance of $1.8 \pm 0.5$~kpc, the implied transverse speed is $235 \pm 45$~km~s$^{-1}$.

Because the transverse speeds inferred from proper motion and scintillation depend differently on the distance, we may use these measurements to calculate a distance consistent with both measurements. Assuming a uniform scattering medium, this distance is 3.6~kpc, with implied transverse speed of  340~km~s$^{-1}$. Not only is this far higher than the transverse speeds measured for any isolated MSPs but would imply a Shklovskii correction greater than the observed $\dot{P}$! We therefore conclude that the scattering medium is not uniform and that the actual distance and speed are closer to 1.8~kpc and 188~km~s$^{-1}$, respectively.

\section{Drifting Subpulse Behaviour}\label{singlep}
Single pulses often consist of several components, called subpulses, which change position, often drifting from pulse to pulse. This phenomenon was first reported by Drake \& Craft (1968)\nocite{dc68} and has since been recognised in roughly 100 radio pulsars \nocite{wes05}(Weltevrede, Edwards \& Stappers~2005). It is usually understood within the {\boldmath$E \times B$} drift model of \cite{rs75} in which radio emission comes from discrete locations, or ``sub-beams'', positioned on a circle around the magnetic pole. The sub-beams rotate and move in and out of our line of sight.  The Ruderman \& Sutherland model has been fairly successful in explaining pulsar drifting patterns. In fact, for PSR~B0943+10, \cite{dr99} have been able to determine that there are 20 sub-beams rotating around this pulsar's magnetic axis, with a rotation timescale close to 100~s.

\begin{figure*}
\includegraphics[angle=0,scale=.65]{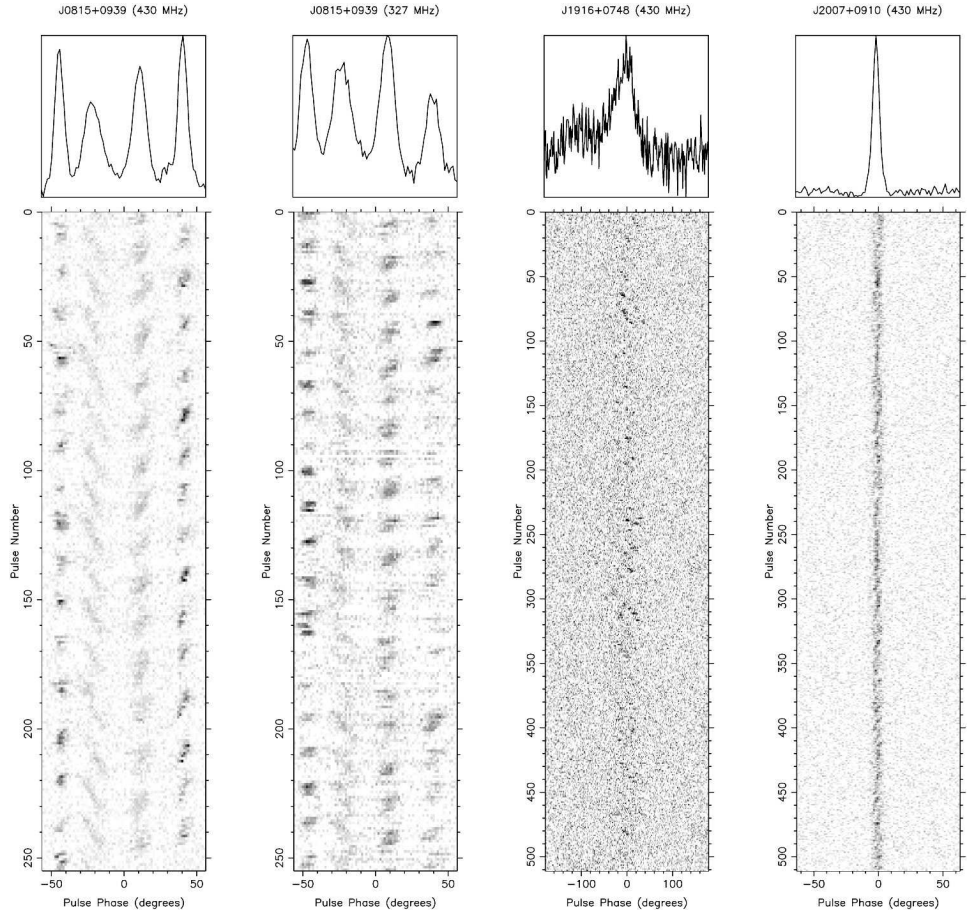}
\caption{The single pulses for pulsars J0815+0939, J1916+0748 and J2007+0912. The intensity is indicated by the linear greyscale (white: low intensity, black: high intensity). The drifting can be seen as diagonal stripes. In PSR~J0815+0939 it is clear that the second component drifts in the opposite sense to the third and fourth. Drifting in the first component is unclear and seems to show some sense reversal.}
\label{SPPlot}
\end{figure*}

In some cases our PSPM search-mode data were of sufficient quality to analyse individual pulses. In our sample three of the pulsars show drifting subpulses. The drifting can be characterised by the drift rate and two periods ($P_2$ and $P_3$; \citealt{Bac73}). The drift rate is simply the longitude of pulse phase drifted by each subpulse per second. The separation in longitude between consecutive subpulses and the separation at a given longitude between successive subpulses are denoted by $P_2$ and $P_3$. The single pulses for the three pulsars are shown in Fig. \ref{SPPlot} and the measured drift rate and $P_3$ for each pulsar are listed in Table \ref{SubPTab}; $P_2$ is not applicable in these cases as there is never more than one subpulse (per component) visible during a single pulse. For PSR J0815+0939, $P_3$ was determined using a fluctuation spectrum analysis, in which a Fourier transform is applied to a series of single pulses (in this case 256) for a range of pulse longitudes.

\begin{table*}
\caption{Parameters obtained from our analyses of the single-pulse observations described in Section \ref{singlep}\label{SubPTab}}
\begin{center}
\begin{tabular}{l l l r@{.}l r@{.}l}
\hline
Name & Frequency (MHz) & Component & \multicolumn{2}{c}{Drift rate (deg~s$^{-1}$)} & \multicolumn{2}{c}{$P_3$(s)}\\
\hline
J0815+0939 & 430 & I   & \multicolumn{2}{c}{-} & 10&87(16)$^a$\\
           &     & II  & $-$1&60(13)           & 10&87(16)$^a$\\
           &     & III & 2&46(19)              & 10&87(16)$^a$\\
           &     & IV  & 1&57(16)              & 10&87(16)$^a$\\
           & 327 & I   & \multicolumn{2}{c}{-} & 8&72(16)$^a$\\
           &     & II  & $-$2&3(5)             & 8&72(16)$^a$\\
           &     & III & 3&2(8)                & 8&72(16)$^a$\\
           &     & IV  & 2&68(18)              & 8&72(16)$^a$\\
J1916+0748 & 430 & I   & 5&4(14)               & 8&6(8)\\
J2007+0912 & 430 & I   & 1&7(6)                & 5&4(7)\\
\hline
\end{tabular}
\end{center}
Figures in parentheses are 1$\sigma$ uncertainties in the least significant digits.\\
These parameters are derived from the highest S/N observation for each pulsar, but the drifting
properties appear to be consistent from epoch to epoch. A more detailed study of the time
evolution of the drifting
properties of J0815+0939 will be presented elsewhere.
a: Determined from fluctuation spectra. Uncertainties are determined by the resolution of
the spectra.
\end{table*}

The single-pulse behaviour of the 645-ms pulsar J0815+0939 is remarkable. This pulsar's profile is unusual, consisting of four components with separations of 41, 58, and 54~ms and not easily described by standard phenomenological classification schemes \cite{ran83}. As shown in Fig. \ref{SPPlot}, drifting is observed within all four components. We refer to these (from left to right) as components I, II, III and IV. The drift within component I is difficult to measure and appears to change sense within an observation. The drift rates of components II, III and IV are very stable from epoch to epoch and do not appear to change sense. However, the drift sense of component II differs from those of components III and IV. While of opposite sign, the drift rates of components II and IV are equal within the errors, indicating a possible relationship. Measurements at 327 and 430 MHz indicate an evolution of the drift rate with frequency, with a higher drift rate at the lower frequency.

While drift sense reversals within a single component are quite common and are attributed to aliasing effects \citep{ggk+04}, it is rare for different components to show different drift senses (Hankins \& Wolszczan 1987; Weltevrede et al.~2005\nocite{hw87,wes05}). How do we interpret this strange behaviour? It is very difficult to reconcile with the standard Ruderman \& Sutherland model in which sparks travel around a circular region centred on the pulsar's magnetic axis. In this case the four components would probably be associated with two nested hollow-cone beams. Similar drift sense would be expected in components I and IV and also II and III but this is not observed. This pulsar's ``bi-drifting'' behaviour has been geometrically modelled by \cite{qlz+04}. While promising, this model is based on one short integration at one frequency. Edwards \& Stappers (2003)\nocite{es03} interpret the behaviour of
B1918+19 as being due to multiple imaging of subbeam systems. Further studies are planned which will determine if there are any aliasing effects while polarisation measurements will help determine the relationship between the multiple components.

In addition to J0815+0939, both PSR~J1916+0748 and PSR~J2007+0912 show weak drifting as shown in Fig.~\ref{SPPlot}. For J1916+0748, the broad integrated profile can be explained as the superposition of a series of drifting subpulses. Further studies of these pulsars with higher sensitivity and time resolution in future may well be worthwhile.

\section{Conclusions}

We have presented a timing and single-pulse study of a sample of 17 pulsars discovered with the Arecibo telescope. These observations provide new insights into a wide range of neutron star physics. We now briefly summarize the new results obtained from the present study.

Extending the timing baseline of the relativistic binary pulsar J1829+2456 presented by Champion et al.~(2004) has allowed the mass of the system to be further constrained, $M=2.59\pm0.02$~M$_{\odot}$, and an upper limit to be placed on the relativistic time-dilation and gravitational redshift parameter, $\gamma<8$~ms. The measurement of $\dot{P}=5.25\pm0.15\times10^{-20}$ has placed an upper limit on the proper motion, $\mu_T=$20~mas~yr$^{-1}$, (using the Shklovskii effect) and a lower limit on the characteristic age, $\tau\sim12.4$~Gyr. Further timing will allow a measurement of this pulsar's proper motion and a determination of the intrinsic $\dot{P}$. A measurement of the parallax through timing is less likely as the current level of precision.

The proper motion of PSR~J1944+0939 has been measured and revealed the highest transverse speed of any isolated MSP, $V_T = 188 \pm 65$~km~s$^{-1}$. An alternative estimate of the speed using interstellar scintillation, $V_{ISS}=235 \pm 45$~km~s$^{-1}$, demonstrates that the scattering medium along the line of sight is non-uniform. These measurements suggest that the space velocity of J1944+0939 is higher than for any other isolated MSP. Further timing observations will ultimately allow the parallax of this pulsar to be measured which will resolve current uncertainties in the transverse speed estimates.

The drifting observed in PSR~0815+0939 is very unusual and warrants further study. All four components of the pulse profile show drifting subpulse behaviour. Of the three components in which we see no drift sense reversals, one of the components has a different drift sense to the others. This is difficult to explain using the standard Ruderman \& Sutherland model. Further observations, including polarimetry, are required to shed light on this remarkable pulsar and the implications it may have for emission mechanism models.

\section*{Acknowledgements}
The Arecibo observatory, a facility of the National Astronomy and Ionosphere Center, is operated by Cornell University in a co-operative agreement with the National Science Foundation (NSF). We thank Alex Wolszczan for making the PSPM freely available for use at Arecibo. Without this superb instrument, the results presented here would not have been possible. DJC is funded by the Particle Physics and Astrophysics Research Council. DRL is a University Research Fellow funded by the Royal Society. ZA is supported by NASA grant NRA-99-01-LTSA-070. FC is supported by the NSF and NASA.

\bibliographystyle{mn2e}
\bibliography{/psr/tex/bib/journals.bib,/psr/tex/bib/psrrefs.bib,/psr/tex/bib/modrefs.bib,/psr/tex/papers/drift-timing/myrefs.bib}

\label{lastpage}

\end{document}